# General Pseudoadditivity of Composable Entropy Prescribed by Existence of Equilibrium

Sumiyoshi Abe

*College of Science and Technology, Nihon University,*

*Funabashi, Chiba 274-8501, Japan*

The concept of composability states that entropy of the total system composed of independent subsystems is a function of entropies of the subsystems. Here, the most general pseudoadditivity rule for composable entropy is derived based only on existence of equilibrium.

Consider a thermodynamic system composed of two independent subsystems, *A* and *B*, in contact to each other. The concept of "composability" [1] states that total entropy of any kind, $S(A, B)$, is given by a function of the same type of entropies of the subsystems, $S(A)$ and $S(B)$:

$$S(A, B) = f(S(A), S(B)). \tag{1}$$

Here, *f* is a certain bivariate function of the $C^2$ −class and is assumed to be symmetric

$$f(S(A), S(B)) = f(S(B), S(A)). \tag{2}$$

The celebrated Boltzmann-Shannon entropy possesses the additivity property [2]

$$f(S(A), S(B)) = S(A) + S(B). \tag{3}$$

Another example may be supplied by the rule

$$f(S_q(A), S_q(B)) = S_q(A) + S_q(B) + Q(q) S_q(A) S_q(B), \tag{4}$$

where $Q(q)$ is a function of the "entropic eindex", $q$, satisfying $Q(1) = 0$.

$Q(q) = 1 - q$ and $Q(q) = q - 1$ correspond to the Tsallis entropy [1] and the modified Tsallis entropy [3], respectively. In both cases, the derivation of $q$ from unity measures the degree of nonextensivity of the entropies. Clearly, additivity in Eq. (3) is recovered in the limit $q \to 1$. The property in Eq. (4) is referred to here as Tsallis-type pseudoadditivity.

To construct thermodynamics based on entropy, it is necessary to define the state of equilibrium, first. This is essentially relevant to the zeroth law of thermodynamics. In this Letter, we derive the most general form of pseudoadditivity of composable entropy based only on existence of equilibrium.

Consider certain extensive variables, $X(A)$ and $X(B)$, of the subsystems, $A$ and $B$, respectively. $X$ may be, for example, internal energy, system volume or particle number. Then, the equilibrium state [4] characterized by the maximum of the total entropy in Eq. (1) with the fixed total amount, $X(A, B) = X(A) + X(B)$, yields

$$\frac{\partial f(S(A), S(B))}{\partial S(A)} \frac{d S(A)}{d X(A)} = \frac{\partial f(S(A), S(B))}{\partial S(B)} \frac{d S(B)}{d X(B)}. \qquad (5)$$

To establish the zeroth law of thermodynamics, by which physical variable (such as temperature, pressure or chemical potential) common to the subsystems in equilibrium can be identified, it is essential to be able to realize factorization of Eq. (5) in the form

$$F(A) = F(B). \tag{6}$$

This means that the following set of equations should hold:

$$\frac{\partial f(S(A), S(B))}{\partial S(A)} = k(S(A), S(B)) \, g(S(A)) \, h(S(B)), \tag{7}$$

$$\frac{\partial f(S(A), S(B))}{\partial S(B)} = k(S(A), S(B)) \, h(S(A)) \, g(S(B)). \tag{8}$$

Here, $g$ and $h$ are some functions, and $k$ is a differentiable bivariate function. In particular, $h$ has to be differentiable. The function, $k$, does not have the factorized form, in general. The symmetry of $f$ shown in Eq. (2) tells us that $k$ is also symmetric

$$k(S(A), S(B)) = k(S(B), S(A)). \tag{9}$$

Since $f$ is of the $C^2$-class, the integrability condition holds, leading to

$$\frac{\partial k(S(A), S(B))}{\partial S(A)} h(S(A)) \, g(S(B)) + k(S(A), S(B)) \frac{d h(S(A))}{d S(A)} g(S(B))$$
$$= \frac{\partial k(S(A), S(B))}{\partial S(B)} g(S(A)) \, h(S(B)) + k(S(A), S(B)) \, g(S(A)) \frac{d h(S(B))}{d S(B)}. \tag{10}$$

Using Eqs. (7) and (8), we rewrite this equation as follows:

$$\frac{1}{k^2(S(A), S(B))} \frac{\partial k(S(A), S(B))}{\partial S(A)} \frac{\partial f(S(A), S(B))}{\partial S(B)} + \frac{dh(S(A))}{dS(A)} g(S(B))$$

$$= \frac{1}{k^2(S(A), S(B))} \frac{\partial k(S(A), S(B))}{\partial S(B)} \frac{\partial f(S(A), S(B))}{\partial S(A)} + g(S(A)) \frac{dh(S(B))}{dS(B)}. \quad (11)$$

This is an identity and therefore leads to the following equations for the factorized and non-factorized parts:

$$\frac{dh(S(A))}{dS(A)} g(S(B)) = g(S(A)) \frac{dh(S(B))}{dS(B)}, \quad (12)$$

$$\frac{\partial k(S(A), S(B))}{\partial S(A)} \frac{\partial f(S(A), S(B))}{\partial S(B)} = \frac{\partial k(S(A), S(B))}{\partial S(B)} \frac{\partial f(S(A), S(B))}{\partial S(A)}, \quad (13)$$

respectively. The general solution of Eq. (13) is given by

$$k(S(A), S(B)) = G(f(S(A), S(B))), \quad (14)$$

where $G$ is an arbitrary differentiable function.

First, we consider the simplest case when $k$ is a constant function. Without loss of generality, such a constant can be set equal to unity. Then, Eqs. (7) and (8) become

$$\frac{\partial f(S(A), S(B))}{\partial S(A)} = g(S(A))\, h(S(B)), \tag{15}$$

$$\frac{\partial f(S(A), S(B))}{\partial S(B)} = h(S(A))\, g(S(B)), \tag{16}$$

respectively. From Eq. (12), it follows that

$$\frac{1}{g(S(A))} \frac{d h(S(A))}{d S(A)} = \frac{1}{g(S(B))} \frac{d h(S(B))}{d S(B)} \equiv \lambda, \tag{17}$$

where $\lambda$ is a separation constant. Using Eq. (17), we rewrite Eqs. (15) and (16) as

$$\frac{\partial f(S(A), S(B))}{\partial S(A)} = \frac{1}{\lambda} \frac{d h_\lambda(S(A))}{d S(A)} h_\lambda(S(B)), \tag{18}$$

$$\frac{\partial f(S(A), S(B))}{\partial S(B)} = \frac{1}{\lambda} h_\lambda(S(A)) \frac{d h_\lambda(S(B))}{d S(B)}. \tag{19}$$

Here, the case $\lambda = 0$ has to be interpreted as the limit $\lambda \to 0$. Integrating these equations, we find

$$f_\lambda(S(A), S(B)) = \frac{1}{\lambda} h_\lambda(S(A))\, h_\lambda(S(B)) + \text{const}. \tag{20}$$

To have the convergent result in the limit $\lambda \to 0$, without loss of generality, we can set the constant term in Eq. (20) equal to $-1/\lambda$ and simultaneously impose the condition

$$\lim_{\lambda \to 0} h_\lambda(S) = 1, \tag{21}$$

for $\forall S$. Thus, we obtain

$$f_\lambda(S(A), S(B)) = \frac{h_\lambda(S(A)) h_\lambda(S(B)) - 1}{\lambda}. \tag{22}$$

This is the form of the function of composability prescribed by existence of equilibrium.

If both of the subsystems are at the completely ordered states, then $S(A, B) = 0$. This means that

$$h_\lambda(0) = 1, \tag{23}$$

for $\forall \lambda$. On the other hand, if only the subsystem $B$ is at the completely ordered state, then we have

$$S(A, B) = S(A). \tag{24}$$

Therefore, from Eq. (22), we conclude

$$\frac{h_\lambda(S(A)) - 1}{\lambda} = S(A), \qquad (25)$$

or equivalently,

$$h_\lambda(S) = 1 + \lambda S. \qquad (26)$$

With this form, Eq. (21) and (23) are clearly fulfilled. Thus, we find that Eq. (22) is equivalent to Eq. (4) with the identification, $\lambda = Q$. In other words, Tsallis-type pseudoadditivity corresponds to the simplest case when $k$ in Eqs. (7) and (8) is a constant function.

In connection with the above result, we wish to mention that in Ref. [5] a nonextensive generalization of the Shannon-Khinchin set of axioms for the Boltzmann-Shannon entropy is given. The axioms presented there are

[I] $S_q(p_1, p_2, \ldots, p_W)$ is continuous with respect to all its arguments and takes its maximum for the equiprobability distribution $p_i = 1/W$ $(i = 1, 2, \ldots, W)$,

[II] $S_q(A, B) = S_q(A) + S_q(B|A) + (1-q) S_q(A) S_q(B|A)$,

$$[\text{III}] \quad S_q(p_1, p_2, \ldots, p_W, p_{W+1}=0) = S_q(p_1, p_2, \ldots, p_W).$$

Here, $p_i$ $(i=1,2,\ldots,W)$ is the probability of finding the total system in its $i$th state. $S_q(B|A)$ is the conditional nonextensive entropy [5] of the subsystem, $B$, given the sunsystem, $A$. Comparing this set with the Shannon-Khinchin one, the only difference between the two is in [II]. (The Shannon-Khinchin axioms are recovered from [I]-[III] in the limit $q \to 1$.) In a particular case when $A$ and $B$ are independent each other, $S_q(B|A) = S_q(B)$ holds, and accordingly [II] becomes the Tsallis-type pseudoadditivity rule in Eq. (4) with $Q(q) = 1-q$. The uniqueness theorem proved in Ref. [5] states that a quantity $S_q$ satisfying [I]-[III] is equal to the Tsallis entropy [1]

$$S_q(p_1, p_2, \ldots, p_W) = \frac{1}{1-q}\left[\sum_{i=1}^{W}(p_i)^q - 1\right] \tag{27}$$

with $q > 0$.

Next, let us discuss the general case in Eq. (14) and put

$$G(f) = \frac{1}{\dfrac{d\,H(f)}{d\,f}}, \tag{28}$$

where $H$ is a certain differentiable function. Then, Eq. (22) is now replaced by

$$H(S(A, B)) = \frac{h_\lambda(S(A)) h_\lambda(S(B)) - 1}{\lambda}. \tag{29}$$

Here, we are using the same notation as in Eq. (22), which may not cause any confusions. For the completely ordered subsystems, we have

$$H(0) = \frac{h_\lambda^2(0) - 1}{\lambda}. \tag{30}$$

Also, if only the subsystem $B$ is at the completely ordered state, then we have

$$H(S(A)) = \frac{h_\lambda(S(A)) h_\lambda(0) - 1}{\lambda}, \tag{31}$$

or equivalently,

$$h_\lambda(S(A)) = \frac{1 + \lambda H(S(A))}{h_\lambda(0)}. \tag{32}$$

Therefore, from Eqs. (29) and (32), we find

$$H(S(A, B)) = \frac{H(S(A)) + H(S(B)) + \lambda H(S(A)) H(S(B)) - H(0)}{1 + \lambda H(0)}. \tag{33}$$

In a particular case when $H(0)$ can be taken to be zero, this equation becomes

$$H(S(A, B)) = H(S(A)) + H(S(B)) + \lambda H(S(A)) H(S(B)). \tag{34}$$

Eq. (33) is the most general pseudoadditivity rule for entropy prescribed by existence of equilibrium. Tsallis-type pseudoadditivity is recovered when $H$ is the identity function.

As an example, let us consider the choice: $H(f) = \sqrt{f}$ and $\lambda \to 0$. In this case, Eq. (34) gives $S(A, B) = S(A) + S(B) + 2\sqrt{S(A)S(B)}$. This rule might be relevant to the black hole entropy, which is proportional to the square of the mass of the black hole [6].

In conclusion, we have derived the most general pseudoadditivity rule for composite entropy based only on existence of equilibrium. We have shown how Tsallis-type pseudoadditivity can be obtained as the simplest case.

This work was supported by the Grant-in-Aid for Scientific Research of Japan Society for the Promotion of Science.